\documentclass[9pt,conference]{IEEEtran}
\usepackage{amssymb,amsthm,amsmath,array}
\usepackage{graphicx}
\usepackage[caption=false,font=footnotesize]{subfig}
\usepackage{xspace}
\usepackage[sort&compress, numbers]{natbib}
\usepackage{stmaryrd}
\usepackage{xcolor}
\usepackage{mathtools}
\usepackage{float}
\usepackage{textcomp}

\begin{document}
\title{BSNeRF: Broadband Spectral Neural Radiance Fields for Snapshot Multispectral Light-field Imaging}
% \title{Snapshot Multispectral Light-field Reconstruction via Broadband Spectral Neural Radiance Fields}
\author{\IEEEauthorblockN{
        Erqi Huang\IEEEauthorrefmark{1}\IEEEauthorrefmark{2}, 
        John Restrepo\IEEEauthorrefmark{3}, 
        Xun Cao\IEEEauthorrefmark{2} and 
        Ivo Ihrke\IEEEauthorrefmark{1}
    }
    \IEEEauthorblockA{
        \IEEEauthorrefmark{1} Center for Sensor Systems (ZESS), University of Siegen, Germany.\\
        \IEEEauthorrefmark{2} School of Electronic Science and Engineering, Nanjing University, China.\\
        \IEEEauthorrefmark{3} K\textbar Lens GmbH, Germany.}
}
\maketitle
\begin{abstract}
Snapshot Multispectral Light-field Imaging (SMLI) is an emerging computational imaging technique that captures high-dimensional encoded data ($x,y,z,\theta,\phi,\lambda$) in a single shot using a low-dimensional sensor.
The accuracy of high-dimensional data reconstruction depends on representing the spectrum using neural radiance field models, which requires consideration of broadband spectral decoupling during optimization.
Currently, some SMLI approaches avoid the challenge of model decoupling by either reducing light-throughput or prolonging imaging time. 
In this work, we propose a broadband spectral neural radiance field (BSNeRF) for SMLI systems. 
Experiments show that our model successfully decouples a broadband multiplexed spectrum. 
Consequently, this approach enhances multispectral light-field image reconstruction and further advances plenoptic imaging.
\end{abstract}

\section{Introduction}
SMLI is a novel technique for acquiring high-dimensional multispectral light-field encoded data in a single shot, simultaneously encoding spatial $(x,y,z)$, angular $(\theta,\phi)$, and spectral $(\lambda)$ information \cite{hua2022ultra}. 
However, high-dimensional data reconstruction is extremely challenging since the problem is underdetermined, making it difficult to decouple and optimize.
Current SMLI systems sacrifice light-throughput by using diffusers or narrow-band filters \cite{manakov2013reconfigurable, zhu2018hyperspectral} or extend imaging time through angular scanning \cite{li2024spectralnerf, li2024spec}. 
Although these approaches simplify some of the reconstruction underdeterminacy, they result in lower signal-to-noise ratios and slower imaging speeds, limiting their practical applications.
Additionally, other studies use a pretrained registration algorithm for multispectral light-field reconstruction \cite{genser2020camera, zhao2017heterogeneous, huang2022high}.
However, it depends on specific datasets, making it inflexible for various scenes.
In this work, we present a high light-throughput SMLI system. 
Based on the characteristics of this system, we propose a self-supervised BSNeRF algorithm for the joint optimization of spatial, angular, and spectral dimensions.

\begin{figure}
\centering
\includegraphics[width=0.48\textwidth]{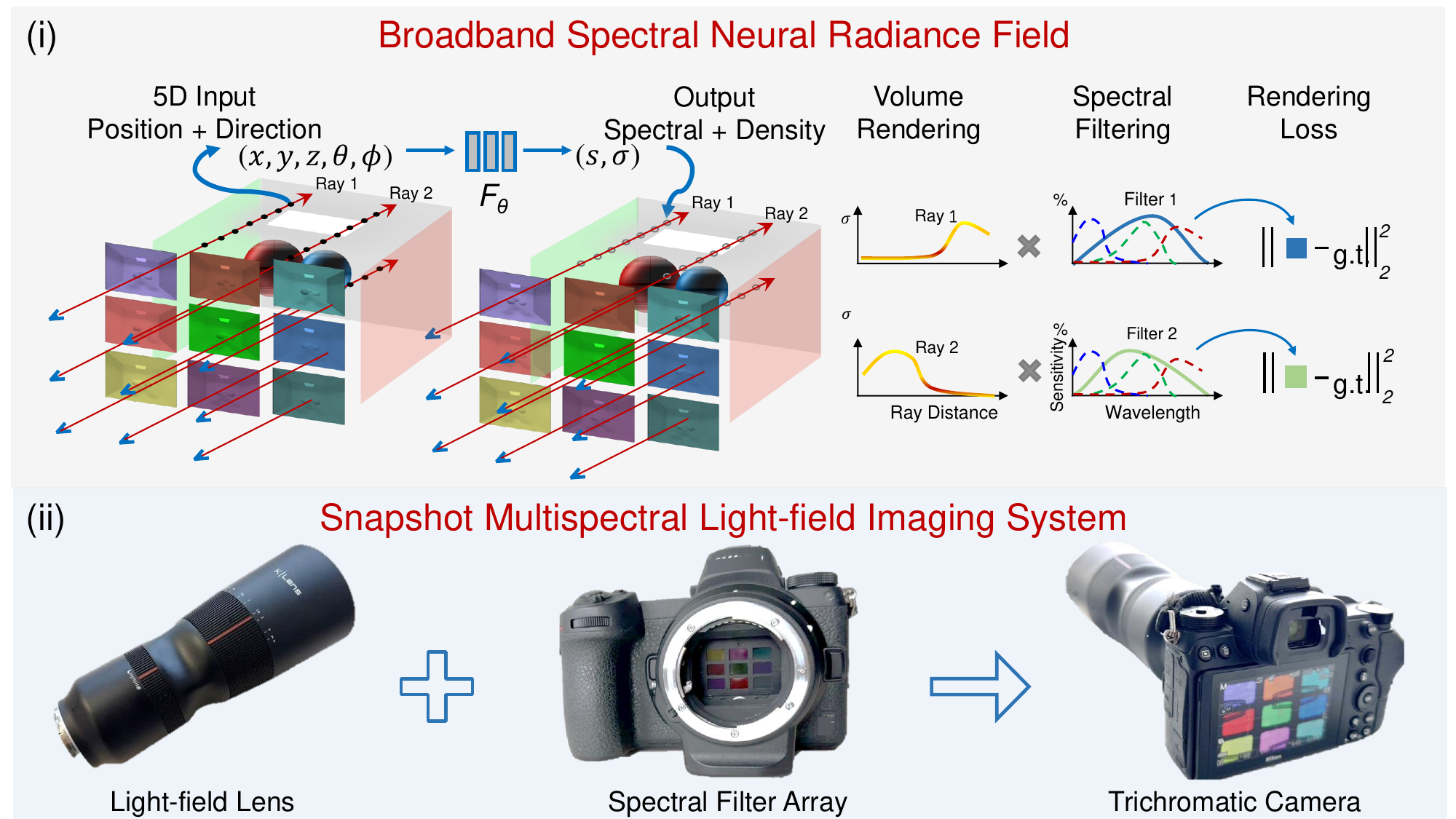}
\caption{(i) Overview of Broadband Spectral Neural Radiance Field: We provide an overview of our broadband spectral neural radiance field scene representation and the differentiable rendering procedure.
(ii) Assembly of Kaleidoscopic SMLI System: We illustrate the assembly of our kaleidoscopic snapshot multispectral light-field imaging system, which incorporates an light-field lens, spectral filter, and trichromatic camera.
}
\label{fig:overview}
\end{figure}

\section{Snapshot Multispectral Light-field Imaging}
\begin{figure}
\includegraphics[width=0.48\textwidth]{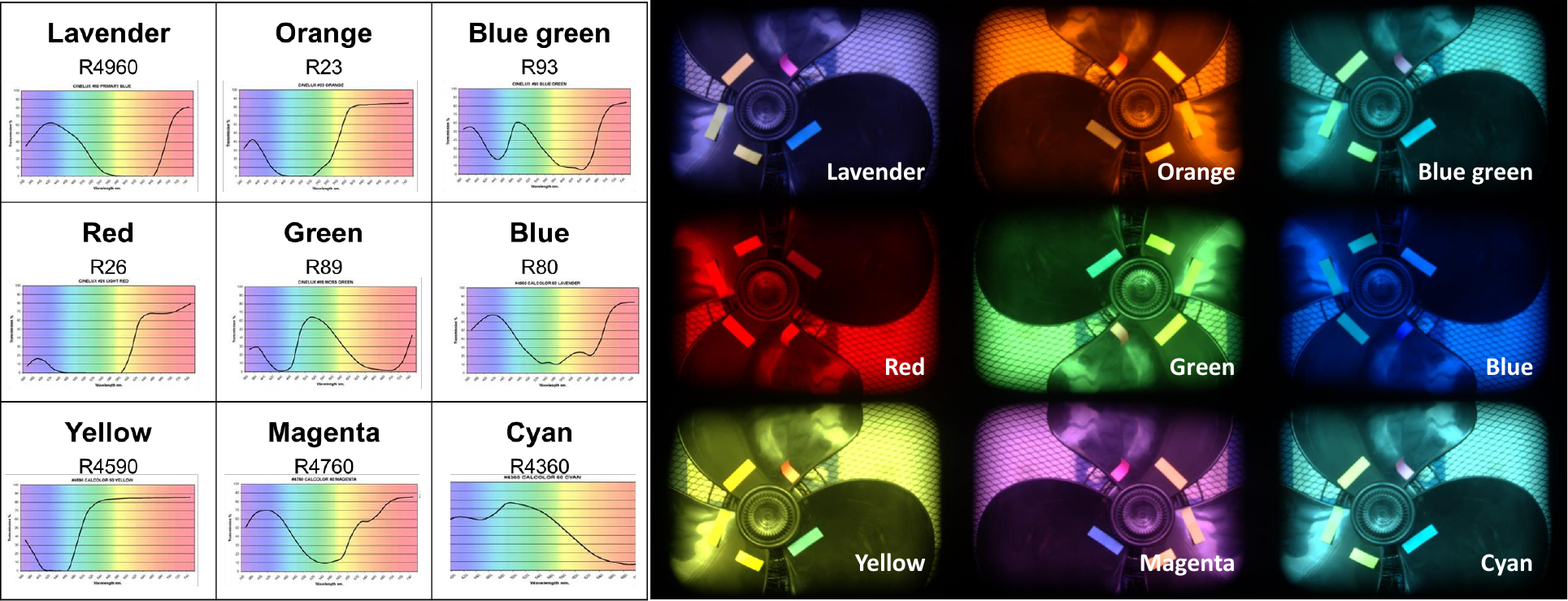}
\caption{
The left panel shows the spectral transmission curves for each filter, labeled with a color name.
The right panel displays subviews corresponding to different filters, including Lavender, Orange, Blue Green, Red, Green, Blue, Yellow, Magenta, and Cyan taken with our prototype.
}
\label{fig:filters}
\end{figure}

We present a kaleidoscopic SMLI system, illustrated in Fig.~\ref{fig:overview} (ii) that provides independent spatial, angular, and spectral encoding across an individual aperture. Our system features a 3$\times$3 array configuration, integrating light-field lens\footnote{K\textbar Lens: https://www.k-lens.de/}, and commercial broadband spectral filters\footnote{Roscolux: https://emea.rosco.com/en/products/catalog/roscolux}, illustrated in Fig.~\ref{fig:filters}, with a single trichromatic SLR camera.
The formulation of multispectral light-field imaging, which incorporates the spatial directions and spectral signals \cite{healey2002radiometric}, is as follows:

\begin{equation}
I_{d,k}(p)=\int_{\Omega} s(p, \lambda) f^{sensor}_{k}(\lambda) f^{filter}_{d}(\lambda) d\lambda
\end{equation}

where $I_{d, k}(p)$ is the intensity of pixel $p$, $k\in \{r,g,b\}$ is the band index of the image, $d$ is the view/filter index, $\Omega = [430nm, 670nm]$ is the visible spectral bandwidth, $s(p, \lambda)$ is the spectral intensity of pixel, $f^{sensor}_{k}(\lambda)$ is the sensor sensitivity of the $k$-th band, and $f^{filter}_{d}(\lambda)$ is the transmission curve of the broadband filter.
Given that our system employs $D$$=$9 spectral transmission filters and utilizes a trichromatic camera with $K$$=$3 color channels, we can acquire broadband spectrally multiplexed images with $K\times D$ channels.

\section{Broadband Spectral Neural Radiance Fields}
To register multispectral images, it is essential to estimate the intrinsic and extrinsic parameters of the camera during model training \cite{wang2021nerf}, because we assume an uncalibrated setting. 
These parameters include the focal length $f$, the rotation matrix $\mathbf{R}$, and the translation vector $\mathbf{t}$.
The camera parameter $\pi$ used for volume rendering consists of $\mathbf{R}$ and $\mathbf{t}$.
Among these, the matrix $\mathbf{R} \in$ SO(3) is particularly challenging.
To recover it, we introduce the Rodrigues formula as follows:

\newcommand{\crossmat}[1]{\left[{#1}\right]_\times}

\begin{equation}
\mathbf{R}=\mathbf{I}+\frac{\sin (\alpha)}{\alpha} \crossmat{\boldsymbol{\phi}} +\frac{1-\cos (\alpha)}{\alpha^2}\crossmat{\boldsymbol{\phi}}^2
\end{equation}

where $\mathbf{I}$ represents the identity matrix, $\omega$ and $\alpha$ denote a normalized rotation axis and a rotation angle, respectively, 
$\phi:=\alpha\omega, \phi \in \mathbb{R}^3$, and $\crossmat{\cdot}$ is a cross product matrix. 

Based on the traditional NeRF volume rendering approach \cite{mildenhall2021nerf}, we introduce broadband spectral sensitivity curves to render the SMLI process, where the spectral view-dependent appearance is modeled by a continuous function $F_\Theta: (\mathbf{x}, \mathbf{d})\xrightarrow{}(s, \sigma)$, which maps world location $\mathbf{x}=(x,y,z)$ and a ray direction $\mathbf{d}=(\theta, \phi)$ to a spectral intensity $s$, as illustrated in Fig.~\ref{fig:overview} (i). The formulation with the output $(s, \sigma)$ from $F_\Theta$ is as follows:

\begin{equation}
\widehat{I}_{d,k}(p) = \mathcal{R}(p,\pi_d\vert \Theta) = \int_{\lambda_n}^{\lambda_f} f(\lambda) \int_{t_n}^{t_f} T(t) \sigma(r(t)) s(r(t), \mathbf{d}, \lambda) dt d\lambda
\end{equation}

where $\mathcal{R}$ is a differentiable rendering function, $\pi_d$ are the camera parameters of the different views, and $\Theta$ presents model parameters. 
$r(t)=\mathrm{o} + t\mathbf{d}$ denotes the ray, 
$\widehat{I}_{d,k}(p)$ is the rendered intensity of the ray integrated over the spectral range from $\lambda_n$ to $\lambda_f$,
$f(\lambda):= f^{sensor}_{k}(\lambda) f^{filter}_{d}(\lambda)$ is the product of the broadband spectral filter transmission and the sensor sensitivity, 
$T(t)=\exp(-\int_{t_n}^t \sigma(r(m)) dm)$ denotes the accumulated transmittance along the ray from $t_n$ to $t$,
$\sigma(r(t))$ represents the attenuation coefficient along the path, and $s(r(t), \mathbf{d}, \lambda)$ denotes the spectral intensity of light at a point along the path in the direction $\mathbf{d}$ from the sensor.

Generating a multispectral image for every viewpoint from SMLI data is challenging because each sub-viewpoint contains only a single type of spectral information.
To ensure that the generated image exhibits similar color characteristics to the available measured subview image, a color loss has been introduced. 
This loss function provides a simple yet effective means of aligning the mean and standard deviation of the color distributions between the generated and the measured images. 
The color loss $\mathcal{L}_\mathrm{color}$ is defined as follows:

\begin{equation}
\mathcal{L}_\mathrm{color}= \frac{1}{D} \sum_{d=1}^{D} \sum_{k=1}^{K}\|\widehat{\mu}_k-\mu^{d}_{k}\|_2^2 + \sum_{k=1}^{K}\|\widehat{\sigma}_k-\sigma_{k}^{d}\|_2^2
\end{equation}

where $\widehat{\mu}_k$ and $\widehat{\sigma}_k$ are the mean and standard deviation of the $k$-th color channel of the generated image, respectively, where $k\in\{R,G,B\}$.
${\mu}_k$ and ${\sigma}_k$ are the mean and standard deviation of the $k$-th color channel of the target image, respectively.

Then, to estimate the spectral intensity of each ray, the fidelity loss is the total square error between the rendered and the true pixel values (for each color channel separately): 

\begin{equation}
\mathcal{L}_\mathrm{fidelity}=\sum_{d=1}^D \sum_{k=1}^K \|\widehat{I}_{d, k}-I_{d, k}\|_2^2
\end{equation}

where $\mathcal{R}$ denotes the set of all rays. $\widehat{I}_{d, k}$ and $I_{d, k}$ are the predicted and ground-truth RGB colors for ray $r$.

Finally, the model can be trained by minimising the error $\mathcal{L}= \alpha\mathcal{L}_\mathrm{fidelity} + \beta\mathcal{L}_\mathrm{color}$ through color loss and fidelity loss:

\begin{equation}
\Theta^*, \Pi^* = \arg \min _{\Theta, \Pi}\mathcal{L}
\end{equation}

where $\Theta^*$ and $\Pi^*$ refers to the model parameters and the camera parameters estimated by the network respectively, the coefficients $\alpha$ and $\beta$ of loss function $\mathcal{L}$ are both set to 0.5 in practice. 

\section{Experiments}
In this section, we discuss the experimental settings and present the results. 
Real-world images are captured using our proposed SMLI system, illustrated in Fig~\ref{fig:filters}. 
The sensor has a resolution of 8288 $\times$ 5520 in RAW format, and each sub-image has a resolution 245 $\times$ 154 after cropping and sub-sampling.
Model training is conducted on an NVIDIA P100 GPU. 
We implement our framework in PyTorch, following the same architecture as NeRF$-$$-$ \cite{wang2021nerf}. 
Three separate Adam optimizers were used for NeRF$-$$-$, camera poses, and focal lengths, each with an initial learning rate of 0.001. The learning rate for the NeRF$-$$-$ model was decreased every 10 epochs by a factor of 0.9954 (equivalent to a stair-cased exponential decay), while the learning rates for the pose and focal length parameters were decreased every 100 epochs by a factor of 0.9. 
All models were trained for 10,000 epochs.

\begin{figure}
\includegraphics[width=0.48\textwidth]{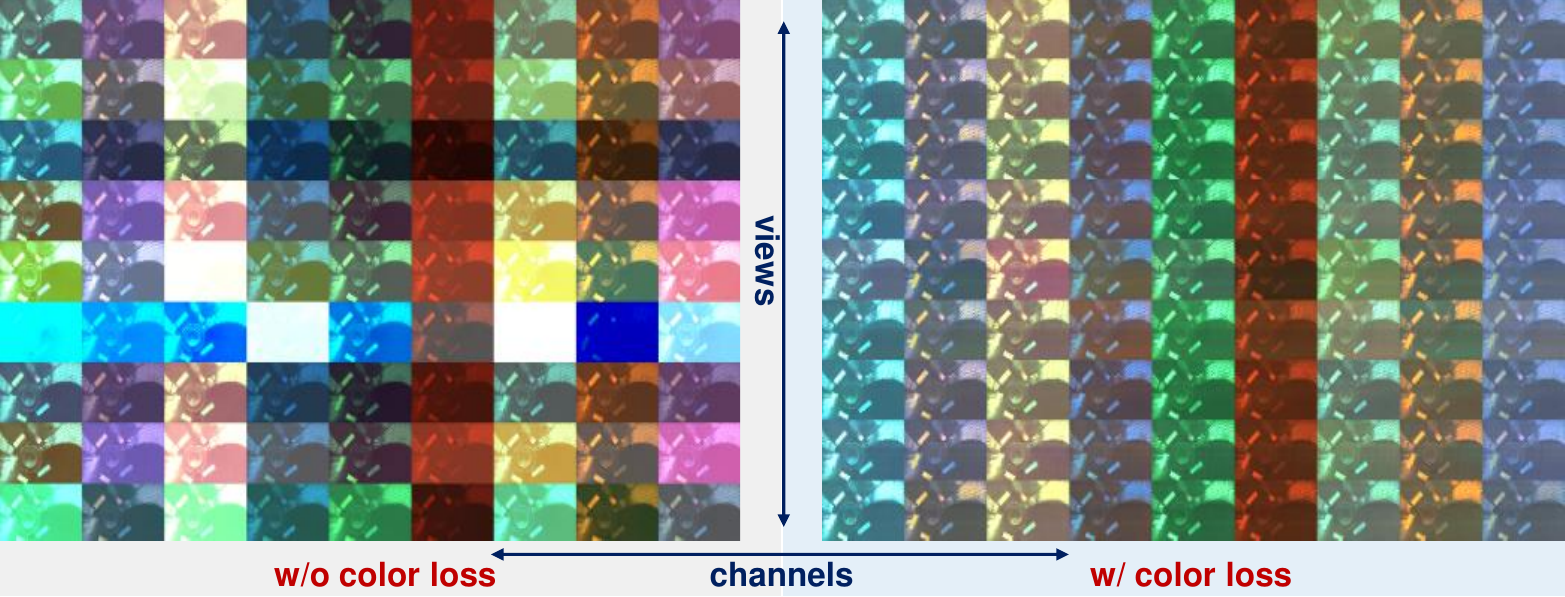}
\caption{Illustration of the reconstruction results of our proposed method applied to snapshot multispectral light-field images captured using our proposed Kaleidoscopic SMLI system.
The left panel shows the reconstructed multispectral light-field images without color loss. In contrast, the right panel presents the reconstructed multispectral light-field images with our proposed additional color loss. It demonstrates the improvement of consistency across views and spectral channels by our proposed color loss.
}
\label{fig:results}
\end{figure}

We present the results of our proposed BSNeRF model applied to the SMLI system. We reconstruct $9\times9$ array of RGB images, corresponding to 9 subviews and 9 spectral filters. There are integrated over 27 channels of spectral intensity generated using BSNeRF with/without color loss, as shown in Fig.~\ref{fig:results}. The fidelity of the reconstruction is evaluated using both color loss and fidelity functions, ensuring that the generated images maintain high spectral accuracy and detailed textural information.

\section{Conclusion}
This work presents BSNeRF, a decoupling model in spatial, angular, and spectral dimensions for a kaleidoscopic SMLI system.
To accommodate arbitrary applications, the camera parameters are jointly optimized during spectral volume rendering.
Leveraging a color statistic prior, our method can generate high-fidelity RGB outcomes in a single snapshot.
Comparative experiments have demonstrated the effectiveness of the proposed model.
In the future, we plan to incorporate the temporal dimension to implement full plenoptic imaging based on the proposed system.

\section{Acknowledgments}
This work was supported by the German Research Foundation
(DFG) under grant FOR 5336 (IH 114/2-1), a CSC Scholarship funded by China Scholarship Council, and a ZESS Scholarship funded by the Lamarr Institute.

\bibliographystyle{IEEEtran}
\bibliography{references}
\end{document}